\documentclass[prb,twocolumn,letterpaper,groupedaddress,amsmath,amssymb,showpacs,aps,english]{revtex4-1}

\usepackage{subfigure}
\usepackage{graphicx}
\usepackage{bm,mathrsfs}
\usepackage{amsmath}
\usepackage{amssymb}
\usepackage{esint}
\usepackage{setspace}
\usepackage{hyperref}

\makeatletter

\begin{document}

\title{Toward engineered quantum many-body phonon systems}
\author{\"O. O. Soykal}
\author{Charles Tahan}
\affiliation{Laboratory for Physical Sciences, 8050 Greenmead Dr, College Park, MD 20740}
\begin{abstract}

Arrays of coupled phonon cavities each including an impurity qubit in silicon are considered. We study experimentally feasible architectures that can exhibit quantum many-body phase transitions of phonons, e.g. Mott insulator and superfluid states, due to a strong phonon-phonon interaction (which is mediated by the impurity qubit-cavity phonon coupling). We investigate closed equilibrium systems as well as driven dissipative non-equilibrium systems at zero and non-zero temperatures. Our results indicate that quantum many-body phonon systems are achievable both in on-chip nanomechanical systems in silicon and distributed Bragg reflector phonon cavity heterostructures in silicon-germanium. Temperature and driving field are shown to play a critical role in achieving these phonon superfluid and insulator states, results that are also applicable to polariton systems. Experimental procedures to detect these states are also given. Cavity-phoniton systems enable strong phonon-phonon interactions as well as offering long wavelengths for forming extended quantum states; they may have some advantage in forming truly quantum many-body mechanical states as compared to other optomechanical systems.

\end{abstract}


\maketitle

\section{Introduction}

Polaritons in coupled-cavity arrays have received great interest for studies of strong correlations and collective behavior in light-matter systems.\cite{Hartmann2008,Panzarini1999,Koch,Littlewood2008} Simultaneously, nanomechanics and optomechanics are driving toward the truly quantum regime of mechanical systems,\cite{Lehnert2011,Painter2011,Girvin2007,Zoller2012,Marquardt2012} where, for example, single photons interact with the lowest mechanical mode of a resonator. It is natural to consider whether these latter systems could exhibit many-body quantum interactions in new configurations, allowing for quantum many-body {\em mechanical} systems. To provide coherent and strong interaction between mechanical modes in a controlled way requires a non-linearity, however, analogous to a photon blockade.\cite{Rabl2011}

Optomechanical coupling---between photon and mechanical modes---may provide one avenue\cite{Zoller2012,Marquardt2012} to produce quantum many-body mechanical systems. At present, however, optomechanical coupling must improve by a factor of $\sim$ 140 to reach the quantum limit.\cite{Marquardt2012,Painter2012} Also, mechanical resonators, considered as a quantum object, have very short de Broglie wave lengths because of their mass, limiting the potential for extended quantum states. An alternate system to enable strong coupling has been proposed for acoustic phonons,\cite{Soykal,Ruskov} where a cavity phonon hybridizes with a semiconductor two-level system (TLS) providing a true analog to the cavity-polariton dubbed a cavity-phoniton, which can easily enter the strong coupling regime. In addition, bulk-like single phonons in silicon can have long thermal de Broglie wave lengths, enabling extended quantum states.

\begin{figure}[ht]
  \centering
  \includegraphics[width=8.5 cm]{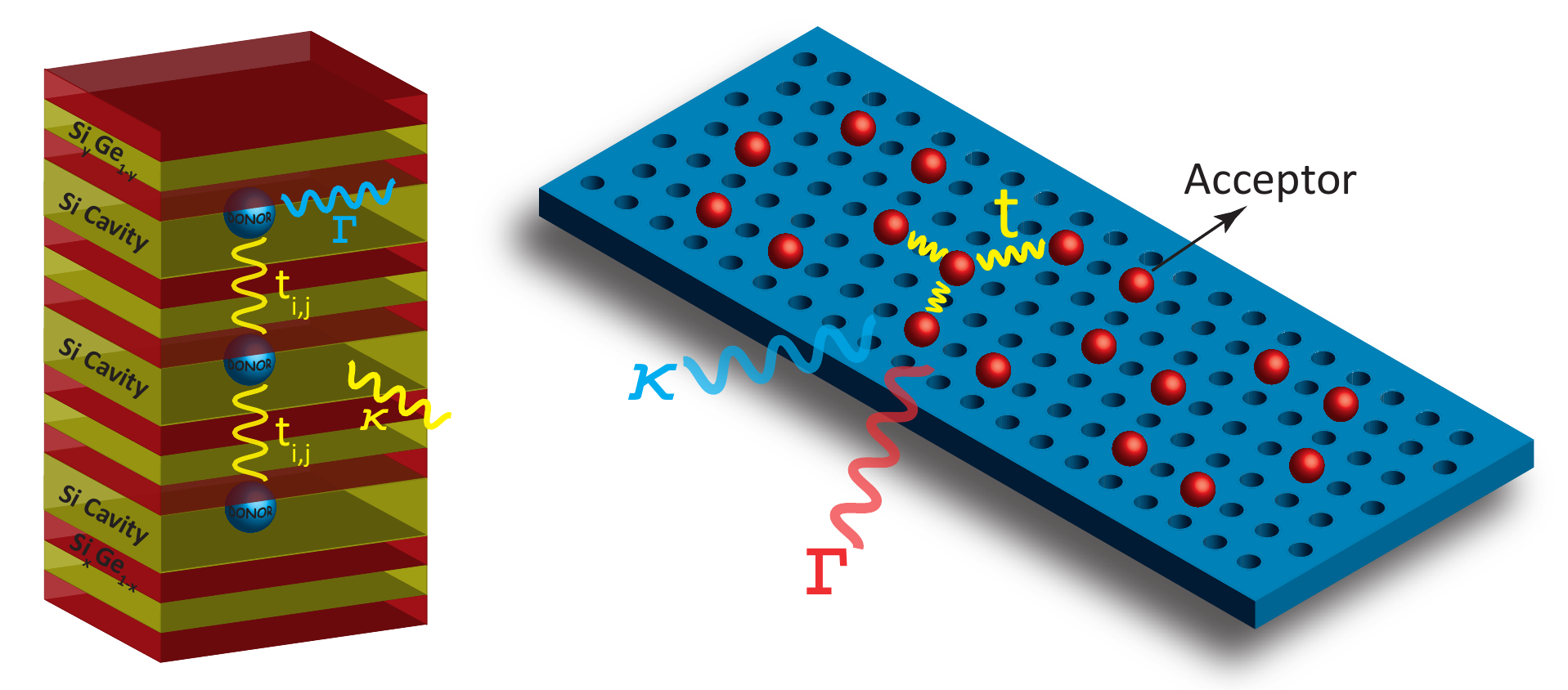}
  \caption{(Color online) Schematic of a strain-matched silicon superlattice heterostructure (acoustic DBR with layers of Si$_x$Ge$_{(1-x)}$/Si$_y$Ge$_{(1-y)}$) consisting of multiple Si cavities, each trapping a single phonon mode,
is shown (left). Every Si cavity site with a loss rate $\kappa$ contains a single donor, acting as a two-level system (TLS) with energy splitting of $\varepsilon$ and relaxation rate of $\Gamma$, strongly coupled to a single cavity phonon mode $\omega$ as well as to each other through an inter-cavity phonon hopping frequency of $t$. A similar two-dimensional phononic crystal structure with acceptors placed at the cavity sites is also shown (right).}
\label{Schematic}
\end{figure}

In this paper, we introduce two experimentally feasible systems in which man-made many-body phonon states can be realized. We begin by identifying the physical parameter regime in which many-body Jaynes-Cummings-Hubbard Hamiltonians\cite{Greentree,Hartman,Makin2008} are realizable and finding such phases as, e.g., the Mott insulator states (``Mott lobes"). Then, as a starting point for considering real experimental setups, we consider a finite array consisting of only two cavity-TLS sites, calculating the super-splitting, the phonon blockade effect, and the response to the driving field strength which would be seen in a measurement. We conclude by considering larger system sizes, showing that extended arrays behave fundamentally differently than the small two-site model under the same hopping and  driving field conditions. 

Schematics of two possible realistic device designs are shown in Fig.~1. Our first device proposal involves the acoustic phonon cavities constructed from distributed Bragg reflector (DBR) heterostructures via alternating layers of Si$_{x}$Ge$_{1-x}$.\cite{Kent,Ezzahri,Trigo} These structures can be further engineered to possess multiple Si cavity regions in a row. In such a setup, the overall reflectivity of the layers between any two Si cavities simply relates to the phonon inter-cavity hopping frequency $t_{ij}$. A suitable donor placed in each of these Si cavities can be strongly coupled (a regime where coupling frequency is much larger than the donor relaxation and cavity loss rates, $g\gg\Gamma,\kappa$) to a specifically chosen single cavity-phonon mode $\omega$.\cite{Soykal} Our second device design is directly borrowed from the concept of nano opto-mechanical phononic crystals. An engineered disturbance in a periodic array of holes can be used for trapping a desired phonon mode in a given region. Placement of an acceptor impurity into each of these regions\cite{Ruskov} will lead to cavity-phonitons with engineered inter-cavity tunneling.

\section{Equilibrium, Grand Canonical}

To determine the parameter range for hopping and transition frequencies of quantum phase transitions, we first consider an equilibrium system in which the phoniton number density is fixed. This is a good approximation when the phoniton lifetime is longer than the thermalization time. For arrays consisting of phosphorus donors (or boron acceptors) and phonons in a silicon phononic crystal or a DBR array (see Fig.~1), the total many-body Hamiltonian is given by the now standard Jaynes-Cummings-Hubbard (JCH) model,\cite{Greentree,Fisher,Koch,Angelakis,Hartman,Makin2008}
\begin{align}
&\mathcal{H}_{JCH}=\mathcal{H}_{JC}-\sum_{\langle i,j\rangle}t_{ij}a_i^\dagger a_j,\label{1}\\
&\mathcal{H}_{JC}=\sum_{i}\left[\varepsilon\sigma_i^+\sigma_i^-+\omega a_i^\dagger a_i+g\left(\sigma_i^+ a_i+\sigma_i^- a_i^\dagger\right)\right],\label{2}
\end{align}
where $a_i (a_i^\dagger)$ is the phonon annihilation (creation) operator at a given cavity site $i$, whereas $\sigma_i^+ (\sigma_i^-)$ is the excitation (de-excitation) operator of the donor at that site. The inter-cavity phonon tunneling is given by the hopping frequency $t_{ij}$ for the nearest neighbor cavity sites $i$ and $j$. The regular Jaynes-Cummings Hamiltonian $\mathcal{H}_{JC}$ corresponds to the interaction of a single mode of the cavity phonon with a TLS.\cite{Milburn} The fast oscillating terms (i.e. $\sigma_i^+a_i^\dagger$) responsible for virtual transitions have been dropped via rotating wave approximation. The third term in Eq.~(\ref{2}) is solely responsible for an effective, non-linear on-site phonon repulsion in analogy with photon blockade.\cite{Kimble,Angelakis}
\begin{figure}[tp]
\centering
\mbox{\subfigure{\includegraphics[width=4.2 cm]{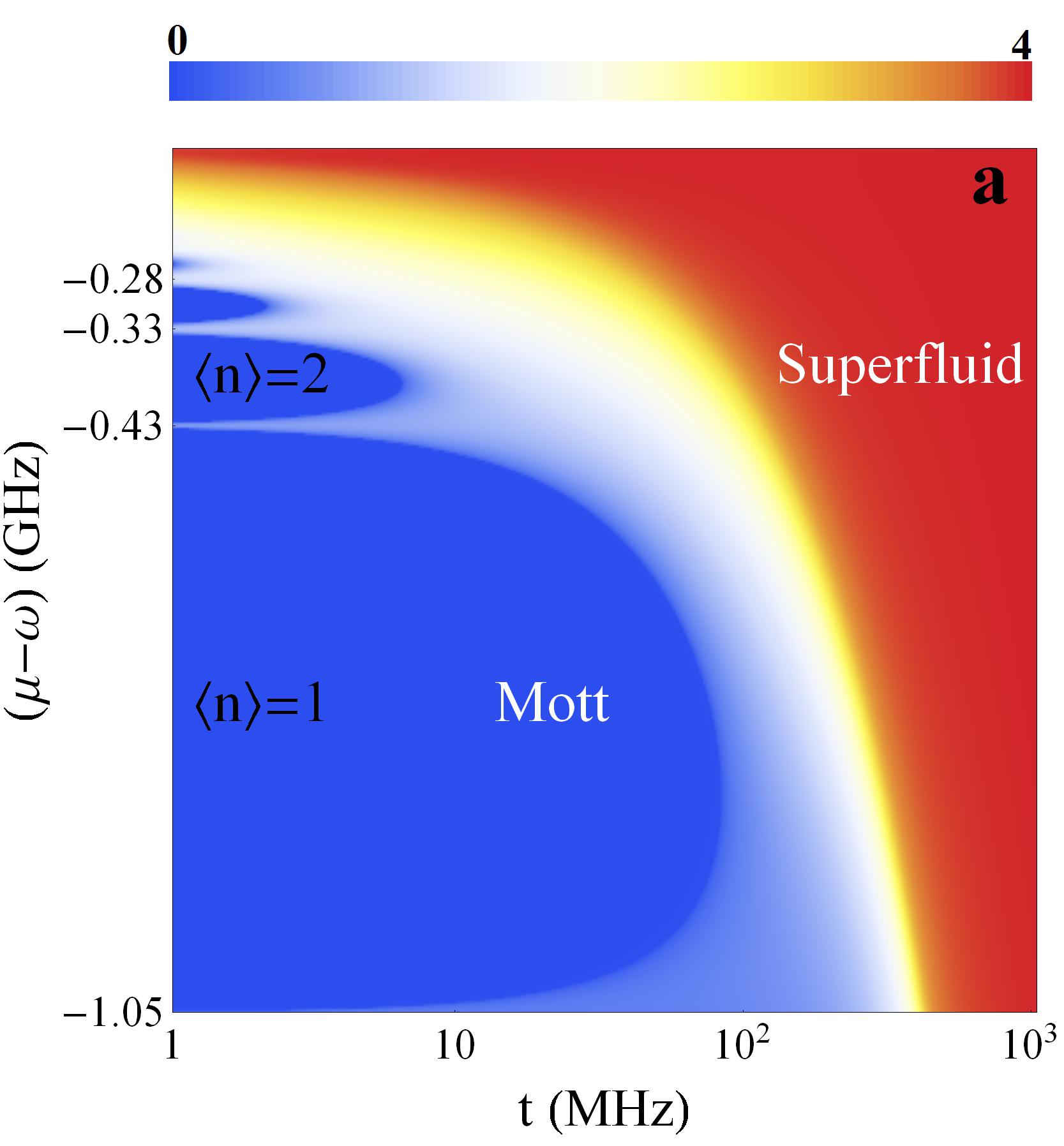}}
\subfigure{\includegraphics[width=4.2 cm]{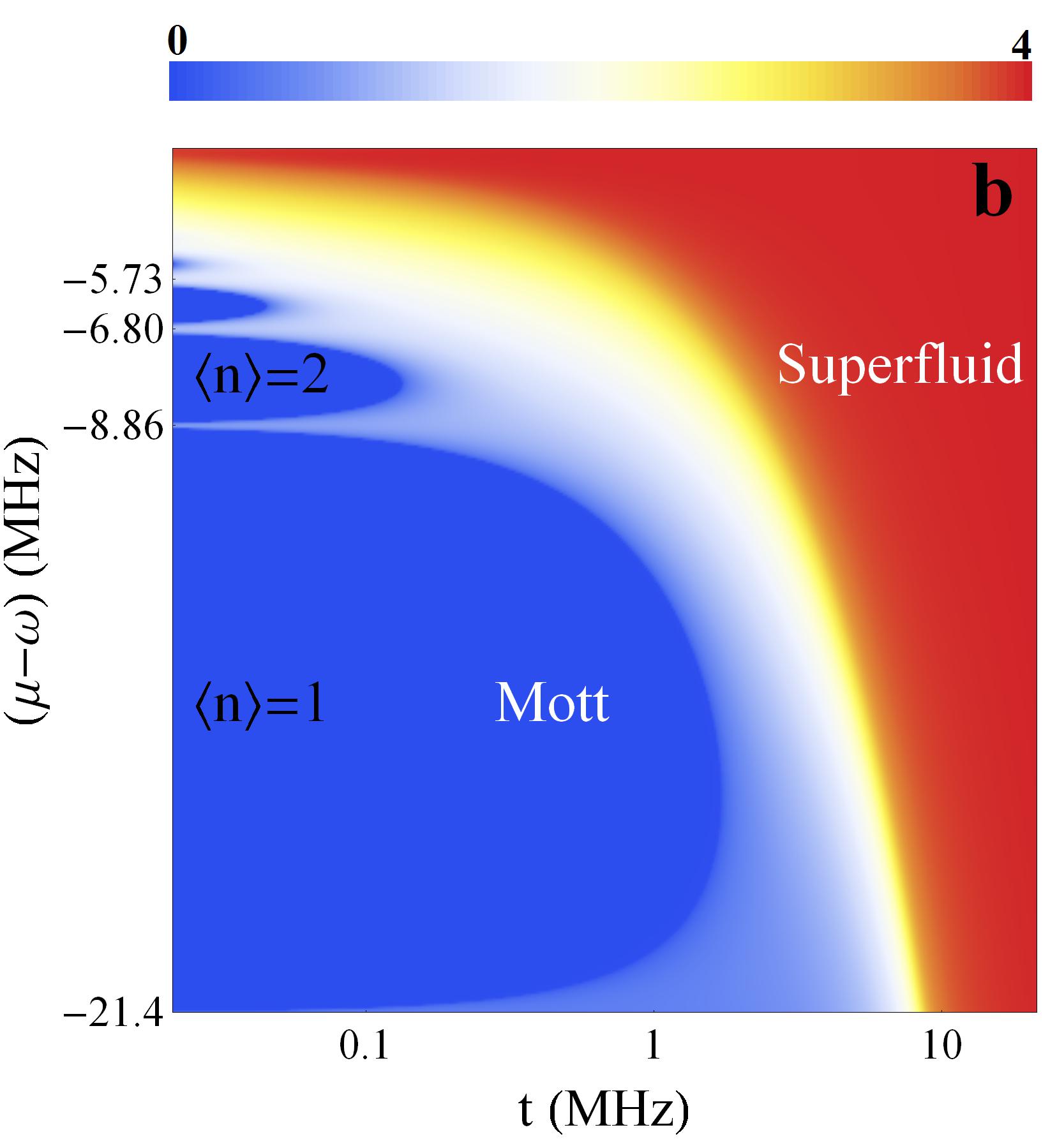} }}\\
\includegraphics*[width=8.5 cm]{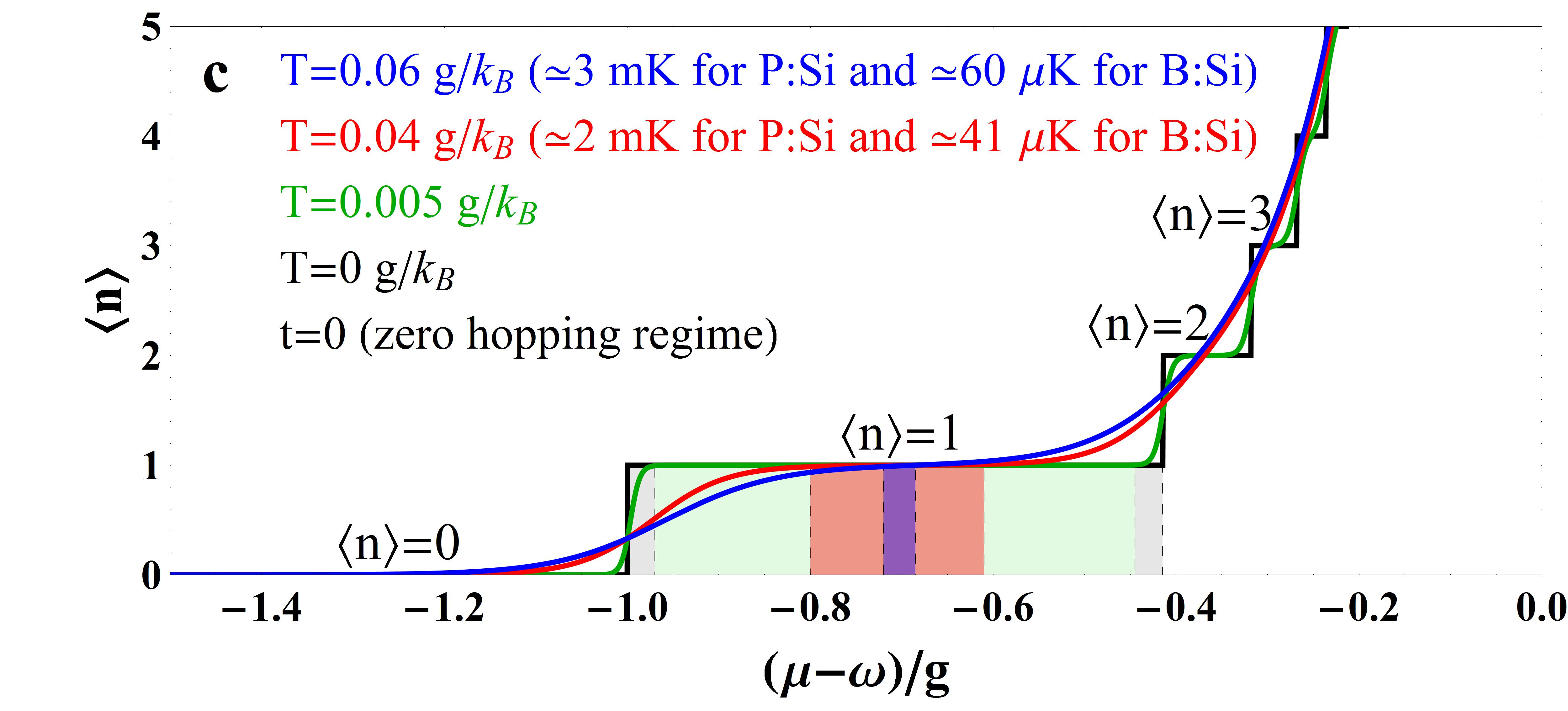}\\
\includegraphics*[width=8.5 cm]{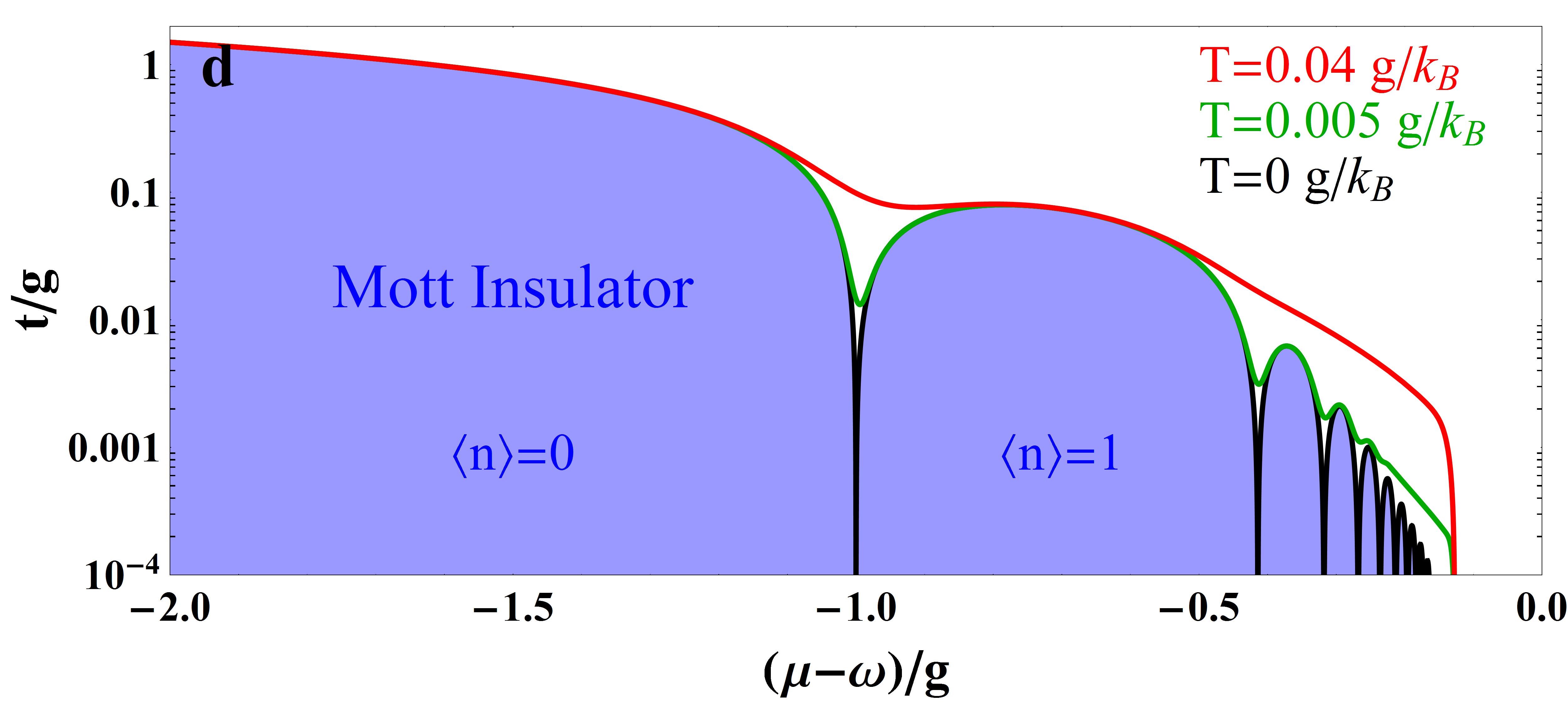} 
\caption{(color online) (a-b) For a many-body phonon-qubit system involving P:Si donors (left) and B:Si acceptors (right), the SF order parameter, $\psi$, is shown as a function of the phonon hopping frequency $t$ and chemical potential $\mu$ with cavity frequency of $\omega$. MI lobes corresponds to the regions of $\psi=0$ (blue) where number of phonons in each lobe is constant ($\langle n\rangle=0,1,2,\dots$). SF phase corresponds to $\psi\neq 0$. (c) Thermal average phonon number per site is shown for various temperatures at zero hopping. Plateaus of constant $\langle n\rangle$ corresponds to MI states. (d) Mott insulator phase boundaries are shown with respect to increasing temperature.} \label{Mottlobes}
\end{figure}
The phase transition between a Mott insulator (MI) and a superfluid phase (SF) can be determined in the grand-canonical ensemble where a chemical potential $\mu$ introduced as $\mathcal{H}=\mathcal{H}_{JCH}-\mu \sum_i N_i$ fixes the number density. The operator $N=\sum_i N_i=\sum_i a_i^\dagger a_i+\sigma_i^+\sigma_i^-$ defines the total number of excitations. For simplicity, one can assume that the random on-site potential with zero mean (e.g. fluctuations of the chemical potential), $\delta\mu_i$, vanishes and $t_{ij}$ is assumed to be a uniform short-range hopping.\cite{Fisher,Koch} In the no-hopping limit, $t_{ij}=0$, each site is occupied by integer number of phonitons $N$ which minimizes the on-site energy $\epsilon(N)=N(\omega-\mu)\pm g\sqrt{N}$ where $\pm$ distinguishes the symmetric and antisymmetric dressed state doublets. However, only the antisymmetric dressed states will be occupied due to their lesser energy. For all values of $\sqrt{N-1}-\sqrt{N}<(\mu-\omega)/g< \sqrt{N}-\sqrt{N+1}$, each site is exactly occupied by $N$ phonitons. Since the number of particles can not be a negative quantity, only $(\mu-\omega)/g<0$ is physically allowed.  If $(\mu-\omega)/g$ is fixed at a value corresponding to $N$ phonitons, i.e. $(\mu-\omega)/g=(\sqrt{N-1}-\sqrt{N+1})/2+\alpha$, the width of $(\mu-\omega)/g$ for a fixed $N$ becomes $\beta=2\sqrt{N}-(\sqrt{N+1}+\sqrt{N-1})$, and the parameter $\alpha$ lies in the range of $-\beta/2<\alpha<\beta/2$. From Fig.~\ref{Mottlobes}a-b, the physical meaning of $\beta$ and $\alpha$ can be readily identified as the $N$ dependent width of each Mott lobe along the $\mu-\omega$ axis and the given distance from the center of each lobe, respectively. Now suppose a weak hopping $t$ is turned on, and it is smaller than the two on-site energies $\delta E_a=g|\beta/2-\alpha|$ and $\delta E_r=g|\beta/2+\alpha|$, required to add or remove one phoniton from the system, respectively. Then the kinetic energy gained by adding a phoniton to the system and allowing it to hop between sites is insufficient to overcome the on-site potential cost. Therefore, for every integer value of  $N$, there lies a finite region in the $t-\mu$ plane in which the number of phonitons is constant at precisely $N$ at each site. Hopping of a phoniton in this region gains a kinetic energy of $t$ while losing a potential energy of $\delta E_a+\delta E_r$. If $t<\delta E_{a}+\delta E_{r}$, as considered here, such hops are energetically not favorable. The hopping probability of a phoniton through $l$ number of sites is roughly $e^{-r/\xi}$, where $\xi\sim 1/\ln\left[(\delta E_{a}+\delta E_{r})/t\right]$. Therefore, regions of constant $N$ of the Mott Lobes corresponds to insulator states wherein the density fluctuations are localized in a linear size of $\xi$ and the compressibility $\partial\epsilon/\partial\mu$ becomes zero, hence, leading to Mott insulating phases. As previously reported,\cite{Koch} in the opposite regime with very large hopping $t/g\gg 1$, the potential term diminishes in comparison to the kinetic term. This yields the degenerate occupation of the lowest localized ground state energy of $\epsilon(N)=N(\omega-\mu)-Nzt$, where the correlation number $z$ is the number of nearest neighbors in a given array geometry. Moreover, if $zt>\omega-\mu$, adding additional phonitons to the system will lower the ground state energy further into the negative values, resulting in an unstable regime. Therefore, the boundary between the MI and the SF phases (Mott lobes) is determined by the value of $\mu$ for which adding or removing a particle does not require any energy. Introducing the SF order parameter, $\psi=\langle a_i\rangle$ via mean-field theory and applying the decoupling approximation, i.e. $a_i^\dagger a_j=\langle a_i^\dagger\rangle a_j+a_i^\dagger\langle a_j\rangle-\langle a_i^\dagger\rangle\langle a_j\rangle$,\cite{Greentree} we obtain the mean-field Hamiltonian,
\begin{equation}
\mathcal{H}_{MF}=\mathcal{H}_{JC}-\sum_i \left\{zt\psi\left(a_i^\dagger+a_i\right)+zt|\psi|^2-\mu N_i\right\}.\label{3}
\end{equation}
Minimization of the ground state energy E of the mean-field Hamiltonian for different parameter ranges of $\mu$, $\omega$, and $t$ for phosphorus (donor) and boron (acceptor) in silicon yields the Mott lobes in Fig.~\ref{Mottlobes}a-b. 

For the calculation of Mott lobes, in the case of phosphorus donor impurity, an acoustic DBR design with correlation number $z=2$ (Fig.~\ref{Schematic}) is used. The donor valley states $1s(A_1)$ and $1s(T_2)$ make up the two-level system with a transition frequency of $\varepsilon=0.7$ THz corresponding to a wavelength of roughly $\lambda\approx 12$ nm.\cite{Soykal} Due to this small wavelength, DBR heterostructures capable of small cavity lengths are the most suitable device structures for maximal coupling. Hence, the large array of silicon/DBR heterostructure phonon cavities can be designed to support a fundamental longitudinal acoustic (LA) phonon mode in resonance with the donor transition ($\omega=\varepsilon$). In the case of the boron acceptor impurity, the transverse acoustic (TA) phonon modes of the cavities are reported to yield the maximum coupling.\cite{Ruskov} TA phonon cavity mode of $\omega=14$ GHz ($\lambda=390$ nm) is needed to be in resonance with the spin splitting (in the presence of a uniform magnetic field of $B=1\,\mathrm{T}$) of the boron valence band acting as a TLS. However, at this large wavelength, DBR phonon cavities are more difficult to construct due to the critical thickness constraint,\cite{Brunner} and 2D phononic crystal designs\cite{Painter2011} need to be implemented. For our calculations, we used a quality factor of $Q=10^5$ currently achievable by both designs. Phonon decay due to anharmonicity and scattering from isotopic point defects have been shown to be smaller than the surface and interface scatterings for both P donor and B acceptor in silicon\cite{Soykal, Ruskov} where the cavity leakage dominates ($\kappa\gg\{\Gamma_{\mathrm{anh}},\Gamma_{\mathrm{imp}}\}$) (See Table I). 

The thermal average phonon number $\langle n\rangle$ per site versus $\mu$ for various temperatures is shown in Fig.~\ref{Mottlobes}c. It is defined by 
\begin{align}
\langle n\rangle&=1/Z_0\sum_{n,\pm}n e^{-E_{n,\pm}/k_B T}\label{4} \\
E_{n,\pm}&=(\omega-\mu)n+(\Delta\pm\sqrt{\Delta^2+4g^2n})/2\label{5}
\end{align}
where $E_{n,\pm}$ are the energy eigenvalues of $\mathcal{H}_{MF}$ with zero hopping and $Z_0=\mathrm{Tr}[e^{-\mathcal{H}_{MF}/k_B T}]$ is the grand canonical partition function for the unperturbed ($t\to 0$) system. The stable MI states (compressibility, $\partial\langle n\rangle/\partial\mu=0$) quickly shrinks with increased temperatures. The maximum temperature allowed to access the first MI state is given as $T=0.04\mbox{--}0.06 \,\,g/k_B$ in terms of coupling strength. 

We also show the temperature dependence of the Mott insulator phase boundaries in Fig.~\ref{Mottlobes}d calculated by an imaginary time evolution formalism of the mean-field Hamiltonian, similar to the Matsubara treatment of temperature.\cite{Mahan} In this formalism, the second term of the mean field Hamiltonian in Eq.~\ref{3} defining the hopping between sites is treated as a pertubation $\mathcal{H}_{t}$ whereas the Jaynes-Cummings term $\mathcal{H}_{JC}$ is assumed to be the homogenous part. Therefore, the grand canonical partition function can be defined as,
\begin{align}
Z&=\mbox{Tr}\left\{\exp{(-\beta \mathcal{H}_{JC})} \hat{T} \exp{\left[-\int_0^\beta d\tau \mathcal{H}_{t}(\tau)\right]}\right\}\nonumber\\
&=\mbox{Tr}\left\{\exp{(-\beta \mathcal{H}_{JC})} U(\beta)\right\},\label{6}
\end{align}
where the imaginary time ($\tau=-it^\prime$) is defined in the range of $0\leq\tau\leq\beta$ ($\beta=1/k_BT$).
$\hat{T}$ and $U(\beta)$ are the time-ordering and imaginary time evolution operators, respectively. Dyson series expansion of the exponential with integral,
\begin{align}
&\exp{\left[-\int_0^\beta d\tau \mathcal{H}_{t}(\tau)\right]}=U(\beta)\label{7}=\sum_{0}^\infty U_n(\beta)\\
&=\sum_{n=0}^\infty\frac{(-1)^n}{n!}\int_0^{\beta}d\tau_1\cdots\int_0^{\beta}d\tau_n\hat{T}\left[\mathcal{H}_{t}(\tau_1)\cdots\mathcal{H}_{t}(\tau_n)\right]\nonumber
\end{align}
yields to the series expansion of the grand partition function $Z=Z_0\left(1+\overline{U_1}(\beta)+\overline{U_2}(\beta)+\cdots\right)$. Thermal average for an operator is given by $\overline{O}=\mbox{Tr}\left\{O\exp{(-\beta\mathcal{H}_{JC})}\right\}/Z_0$ with respect to the unperturbed Hamiltonian, where $Z_0=\mbox{Tr}\left\{\exp{(-\beta \mathcal{H}_{JC})}\right\}$. With this expansion, the definition of free energy $F=-\ln{Z}/\beta$ can be put into the form of $F=-\ln{Z_0}/\beta+\sum_i\alpha\psi^2+o(\psi^4)$ for all sites $i$ with superfluid order parameter $\psi$ upto second order. The solutions for $\alpha=0$ where the symmetry breaking occurs yields the boundaries of the Mott-insulator superfluid phase transition as shown in Fig.~\ref{Mottlobes}d.

\begin{table}
{\footnotesize
\begin{tabular}{|l|c|c|c|}
\hline
Parameter & Symbol & {\scriptsize P:Si}\cite{Soykal} & {\scriptsize B:Si}\cite{Ruskov} \tabularnewline
\hline
Resonance frequency  & $\omega_{\rm {r}}/2\pi$  &  $730\,{\rm {GHz}}$  & $14\,{\rm {GHz}}$ \tabularnewline
\hline
Coupling strength  & $g/2\pi$  & $1\,{\rm GHz}$  & $21.4\,{\rm MHz}$ \tabularnewline
\hline
Wavelength & $\lambda$ & $\sim 12$ nm & $\sim 390$ nm \tabularnewline
\hline
Cavity lifetime & $1/\kappa$ & $ 22\,{\rm ns}$ & $1.14\,{\rm \mu s}$  \tabularnewline
\hline
TLS lifetime  & $1/\Gamma$ & $8.2\,{\rm {ns}}$  & $0.14\,{\rm \mu s}$ \tabularnewline
\hline
{\scriptsize\#} Rabi flops  & ${2g/(\!\kappa\!+\!\Gamma)}$  & $\sim 102$  & $\sim 34$  \tabularnewline
\hline
\end{tabular}
}
\caption{
Parameters used for a cavity phonon-TLS pair consisting of a phosphorus (P) donor or a boron (B) acceptor in silicon.
\label{tab:Key-rates}
}
\label{table:parameters}
\end{table}
\begin{figure}[tp]
\begin{center}
\setlength{\tabcolsep}{0mm}
\begin{tabular}{c}\includegraphics[width=6.5 cm]{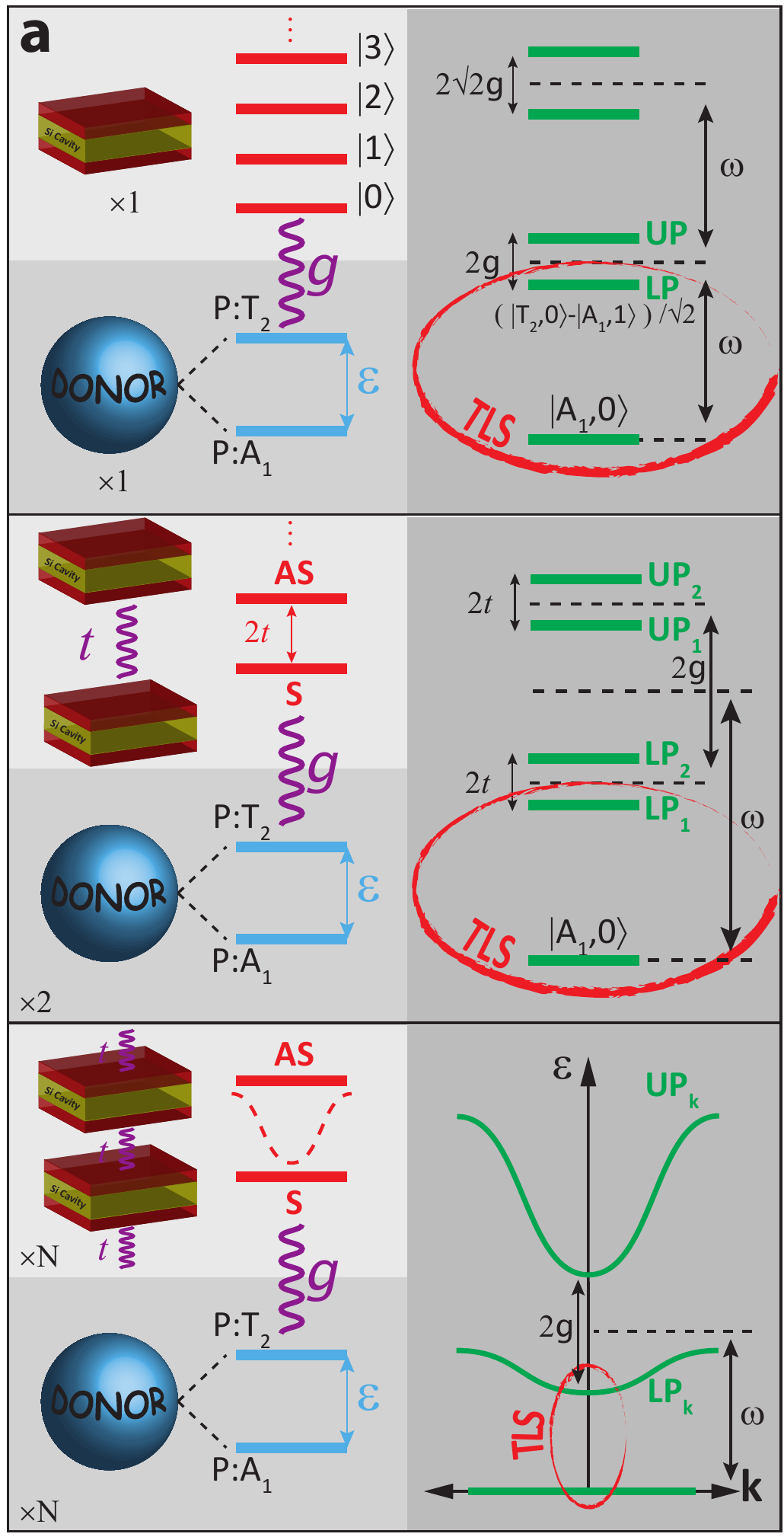}\\
  \includegraphics*[width=8.5 cm]{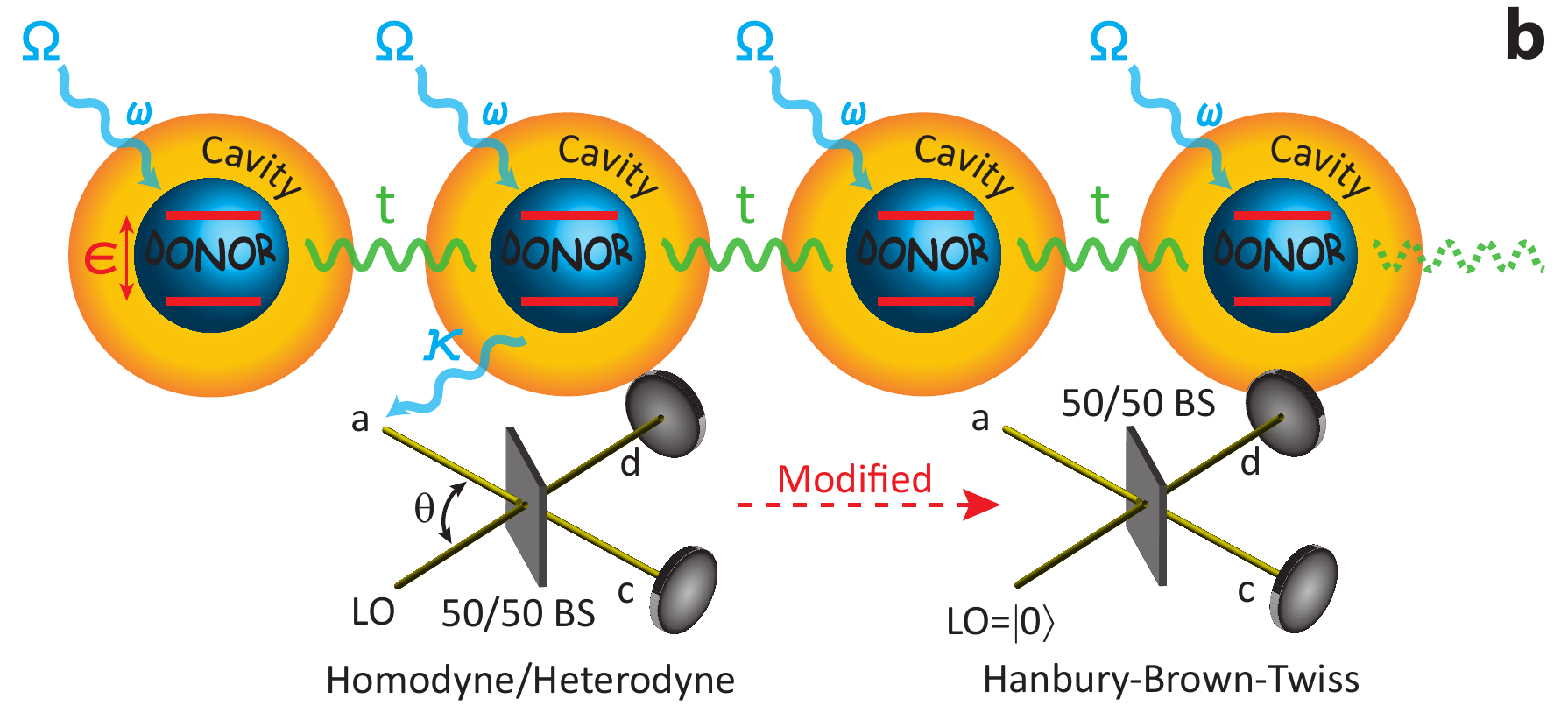} 
	\end{tabular}
\caption{(color online) \textbf{(a)} Energy schematics of a single, two coupled, and an infinite array of coupled cavity phonon-donor pairs are shown. Phosphorus donor energy splitting is $\varepsilon$ between intervalley states $\mathrm{T}_2$ and $\mathrm{A}_1$. Cavity anti-symmetric (AS) and symmetric (S) eigenstates split by the hopping bandwidth $\propto 2t$ are shown for the coupled two and the infinite array systems. Energy diagrams are given in terms of lower phoniton (LP) and upper phoniton (UP) quasi-particle branches in the dressed state representation. \textbf{(b)} Experimental read-out scheme from a single site by a homodyne or heterodyne or modified HBT setup.}\label{Energyplots}
  \end{center}
  \end{figure}
\section{Non-Equilibrium, Driven, Dissipative}

Let us assume that the many-phoniton system is driven at each site by a phonon field of amplitude $\Omega_i$ and frequency $\omega_d$. Switching to the rotating frame of the driven field yields the time-independent Hamiltonian given by
\begin{align}
\mathcal{H}_{S}&=\sum_{i}\left[\Delta\varepsilon\sigma_i^+\sigma_i^-+\Delta\omega a_i^\dagger a_i+g\left(\sigma_i^+ a_i+\sigma_i^- a_i^\dagger\right)\right]\nonumber\\
&-\sum_{\langle i,j\rangle}t_{ij}\left(a_i^\dagger a_j+a_i a_j^\dagger\right)+\sum_i\Omega_i\left( a_i^\dagger + a_i \right),\label{8}
\end{align}
where $\Delta\varepsilon=\varepsilon-\omega_d$ $(\Delta\omega=\omega-\omega_d)$ is the detuning between the driving field and the TLS (cavity).  In the case of dissipation defined by the cavity loss rate ($\kappa$) and the qubit relaxation rate ($\Gamma$), the master equation for the density matrix is given by
\begin{equation}
\dot{\rho}=-i[\mathcal{H}_S,\rho]+\kappa\sum_i L[a_i]\rho+\Gamma\sum_i L[\sigma_i^-]\rho,\label{9}
\end{equation}
where the Lindblad super operator is defined as $L[\hat{O}]\rho=\hat{O}\rho \hat{O}^\dagger-\{\hat{O}^\dagger \hat{O},\rho\}/2$.\cite{LendiAlicki} The number of elements of the density matrix $\rho_{i,j}$ needs to be determined from Eq.~(\ref{9}) is given by $(2(\Lambda+1))^{2n_c}$, where $n_c$ is the number of cavities with a single donor/acceptor inside.

\begin{figure}[tp]
\begin{center}
\setlength{\tabcolsep}{0mm}
\begin{tabular}{c}
  \includegraphics*[width=8.5 cm]{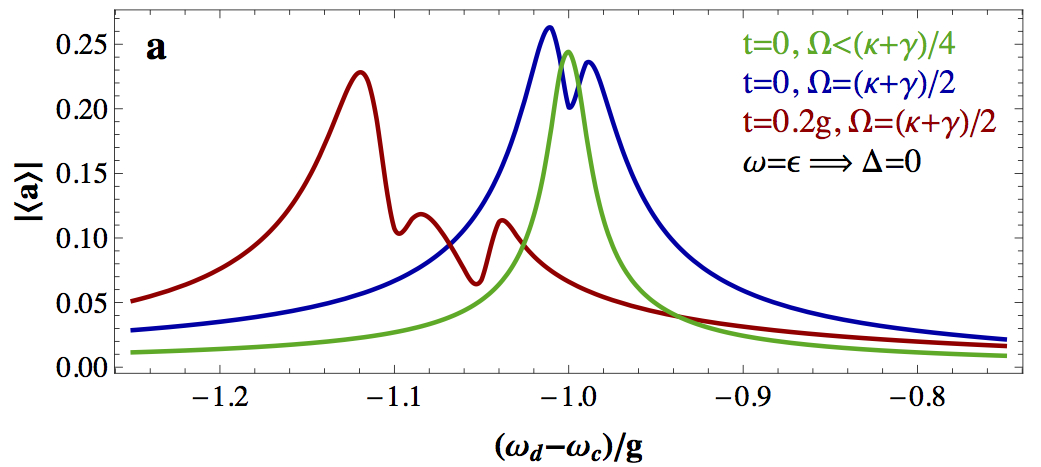} \\
  \includegraphics*[width=8.5 cm]{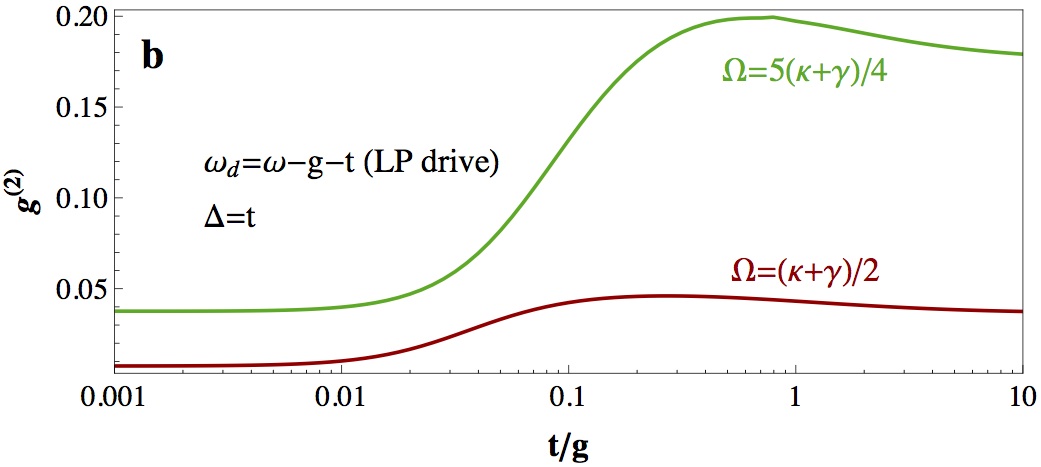} \\
  \includegraphics*[width=8.5 cm]{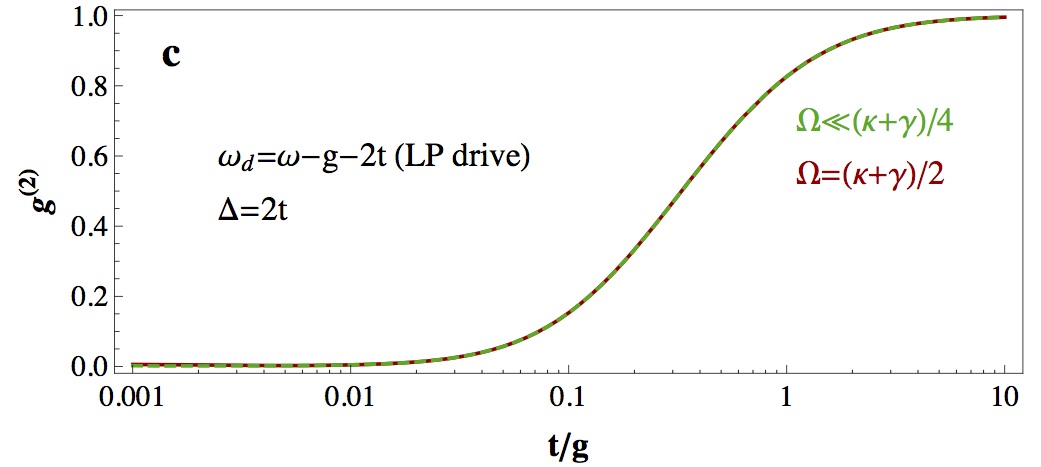}
  \end{tabular}
\caption{(color online) \textbf{(a)} For two coupled cavities each containing a resonant phosphorus P donor, the transmission amplitude versus the detuning between the coherent drive ($\omega_d$) of strength $\Omega=2\Omega_c$ and the cavity field ($\omega$) is shown for hopping frequencies of $t=0,t=0.2g$. For a weak drive ($\Omega<\Omega_c$) and zero hopping $t=0$, system exhibits a Lorentzian response (green line) \textbf{(b)} The second-order coherence function $g^{(2)}$ versus the hopping frequency for drive strengths of $\Omega=2\Omega_c$ and $\Omega=5\Omega_c$, both in resonance with the LP branch ($\omega_d=\omega-g-t$). Donors are detuned by the hopping bandwidth $\Delta=t$ and in resonance with the symmetric cavity-phonon mode. \textbf{(c)} For an infinite array, $g^{(2)}$ versus hopping for $\Omega\ll\Omega_c$ and $\Omega=2\Omega_c$. Donors detuned by $\Delta=2t$.}\label{g2plots}
  \end{center}
  \end{figure}

First, we examine the single phoniton system under different driving field and hopping conditions. This can be done by driving and measuring the heterodyne amplitude of a single site in the case of zero hopping ($t=0$) and resonance ($\varepsilon=\omega$). As seen from the Fig.~\ref{g2plots}a (green line), for weak driving field strengths smaller than the critical value $\Omega<\Omega_c=(\kappa+\Gamma)/4$,\cite{Bishop} the system initially lies in the linear response regime and it exhibits a Lorentzian response to the driving field frequency. The critical coherent drive strength is estimated as $\Omega_c^P\sim 42$ MHz or $\Omega_c^B\sim 2$ MHz for a phoniton composed of phosphorous donor or boron acceptor, respectively. With increasing field strengths, this response breaks down and a super-splitting\cite{Bishop} of the phonon field amplitude occurs (blue line). Intuitively, this behavior can be understood as a coupling of the driving field only with the antisymmetric 1$^{st}$ dressed state ($(|0,e\rangle-|1,g\rangle)/\sqrt{2}$) and the ground state $|0,g\rangle$, therefore forming a two-level system (TLS) as shown in the top row of Fig.~\ref{Energyplots}a. TLS treatment will stay valid with the driving field strength as long as the non-linearity of the Jaynes-Cummings Hamiltonian will only allow single-phonon excitations, preventing access to the higher multiple excitation manifolds and therefore causing a phonon-blockade. In a single cavity system, the lowest two and single excitation energies are given by  $\epsilon_2=2\omega-g\sqrt{2}$ and $\epsilon_1=\omega-g$, respectively. This yields to the necessary condition $\Omega_i\ll g(2-\sqrt{2})$ ($\Omega_i\ll\epsilon_2-2\epsilon_1$) of single-excitation-only subspaces of the system, also known as the ``dressing of the dressed states".\cite{Carmichael,Nunnenkamp} As the single phoniton system still exhibits super-splitting ($\Omega=(\kappa+\Gamma)/2$), turning on the hopping parameter $(t =0.2 g)$ makes the two phoniton states (one phoniton in each cavity) available to occupation. This results with a clear shift in eigenfrequencies and an appearance of a third peak (red line) at the heterodyne amplitude spectrum.

\subsection{Measurement} 
The MI and the SF states exhibit different coherence characteristics which can be accessed via coherence (correlation) function measurements\cite{Tomadin,Davidovich} in setups similar to a modified homodyne or heterodyne setup or a Hanbury-Brown-Twiss setup.\cite{Milburn} Each of these techniques generally requires single-phonon detectors. However, even with single-phonon detectors unavailable, another useful tool, a so called phonon-to-photon translator (PPT),\cite{Painter} can be deployed to coherently convert phonons to photons, therefore allowing the optical detection techniques to be applied on the cavity phonon TLS if necessary (see Fig.\ref{Energyplots}b). The zero-time delay second order coherence function is defined by $g^{(2)}(0)=\langle a^\dagger a^\dagger a a\rangle/\langle a^\dagger a\rangle^2=\left(\langle(\Delta n)^2\rangle-\langle n\rangle\right)/\langle n\rangle^2+1$, where the variance is $\Delta n=n-\langle n\rangle$. The MI phase corresponds to a constant phonon number with zero variance $\Delta n=0$; hence, identified by $g^{(2)}(0)=1-1/\langle n\rangle<1$. On the other hand, the SF phase possesses a constant phase with fluctuating phonon numbers and is represented by a coherent state. Using the definition for coherent states $a|\alpha\rangle=\alpha|\alpha\rangle$, the correlation function for SF state yields to $g^{(2)}(0)=1$.\cite{Glauber} 

For the two-coupled-phoniton case, we calculated the second-order coherence function $g^{(2)}$ versus the hopping frequency for different field strengths. Through out all hopping frequencies, qubits were kept detuned from their encapsulating cavity mode by $\Delta=\omega-\varepsilon=t$ to ensure a resonance with the symmetric mode (lowest) of the overall coupled cavity mode. An energy level schematic for this configuration is shown in the middle row of Fig.~\ref{Energyplots}a. At this detuning choice, the eigenenergy difference between the ground state (GS) and the lower phoniton (LP) branch is given by a simple relation $\Delta E=\omega-g-t$. The driving field always kept in resonance with this energy difference $\omega_d=\Delta E$ to simulate a TLS system. However, for resonant driving purposes, this detuning is not necessary, as long as one can determine the energy difference between the GS and the LP each time hopping and/or coupling parameters are changed. As shown in Fig.~\ref{g2plots}b, even in the case of a strong driving field, $\Omega\gg\Omega_{c}$, the two phoniton system exhibits a phonon anti-bunching. 

For large cavity arrays (bottom row of Fig.\ref{Energyplots}a), the mean-field theory and density matrix master equation can be applied together for weak coherent drive and the strong coupling regime.\cite{Nissen,Tomadin} Starting from the Hamiltonian in Eq.~(\ref{8}), application of the mean field $\psi=\langle a\rangle$ and the decoupling approximation yields to
\begin{eqnarray}
\mathcal{H}_{MF}^\prime&=&\sum_{i}\left[\Delta\varepsilon\sigma_i^+\sigma_i^-+\Delta\omega a_i^\dagger a_i+g\left(\sigma_i^+ a_i+\sigma_i^- a_i^\dagger\right)\right.\nonumber\\
&-&\left.zt \left(a_i^\dagger \psi+a_i\psi^*-\psi^2\right)+\Omega_i\left( a_i^\dagger + a_i \right)\right],\label{10}
\end{eqnarray}
in the presence of a coherent driving field. Including the cavity loss and qubit relaxation, the master equation is same as Eq.~(\ref{9}), only with the driven system Hamiltonian $\mathscr{H}_S$ replaced by the mean-field Hamiltonian $\mathscr{H}_{MF}^\prime$. Similar to the two phonon/qubit site (however, now with cooperativity of $z=2$ due to two nearest neigbors for each site), donors are kept detuned by $\Delta=2t$ to be in resonance with the LP branch and driving field applied in resonance with the two-level splitting of $\omega_d=\omega-g-2t$. The SF order parameter $\psi$ is evaluated by the self-consistency check $\psi=\mathrm{Tr}\left(\rho a\right)$. For phonitons composed of P donors, we calculated the second-order coherence function $g^{(2)}$ versus the hopping frequency for two different field strengths, $\Omega\ll \Omega_c$ and $\Omega=2\Omega_c=84\,\mathrm{MHz}$ in Fig.~\ref{g2plots}c. For our particular donor choice, the critical drive strength is much smaller than the coupling strength $\Omega_c/g\sim 0.006$ due to already small amounts of donor relaxation and cavity loss. For a boron B acceptor, the ratio is estimated as $\Omega_c/g\sim 0.094$. The infinite phoniton array exhibited a smooth transition from incoherent to coherent case, as expected, indicating a phase transition from MI to SF state by increasing the hopping frequency.

\begin{figure}[tp]
\begin{center}
\setlength{\tabcolsep}{0mm}
\begin{tabular}{c}
  \includegraphics*[width=8.5 cm]{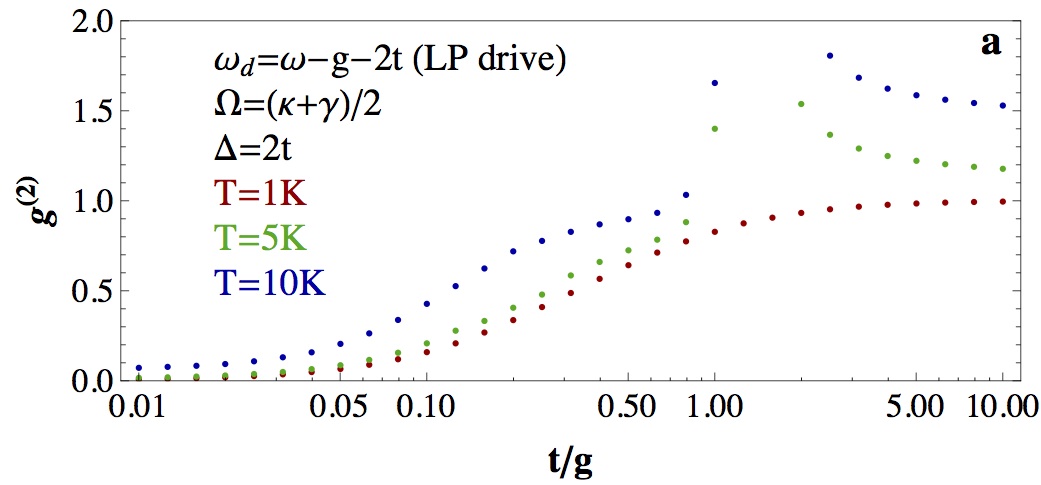}\\
	\includegraphics*[width=8.5 cm]{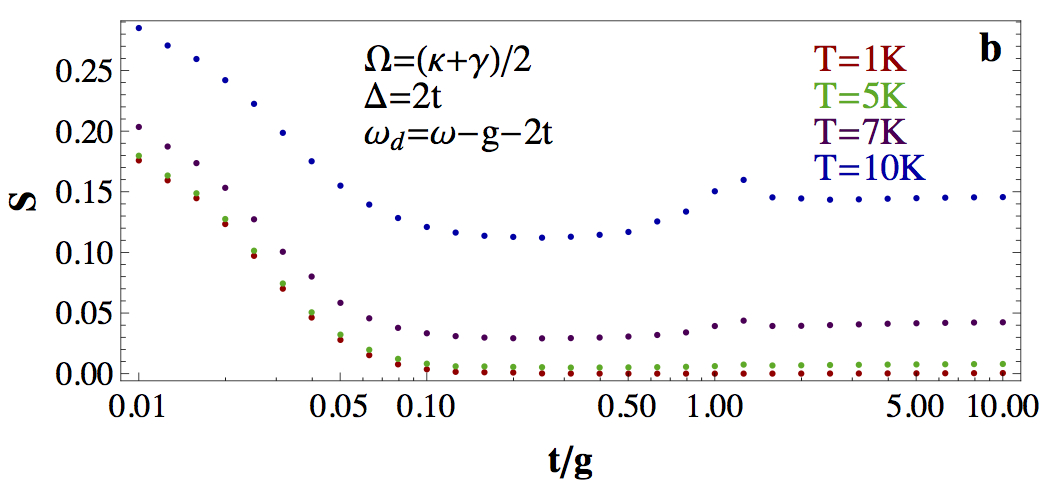}
  \end{tabular}
\caption{(color online) \textbf{(a)} For an infinite array of qubit/phonon cavity sites, the second-order coherence function $g^{(2)}$ versus the hopping frequency for a drive strength of $\Omega=2\Omega_c$, both in resonance with the LP branch ($\omega_d=\omega-g-2t$). Donors are detuned by the hopping bandwidth $\Delta=2t$. \textbf{(b)} Entropy $S$ of the overall system is shown for $T=1\mathrm{K}$, $5\mathrm{K}$, $7\mathrm{K}$, and $10\mathrm{K}$ under the same driving and detuning conditions.}\label{tempcoherence}
  \end{center}
  \end{figure}

In the case of finite temperature for the infinite array of phonon cavity/qubit sites (bottom row of Fig.~\ref{Energyplots}a), the non-equilibrium driven dissipative density matrix includes the cavity phonon field damping due to the coupling to the thermal phonons of the environment, given as
\begin{align}
\dot{\rho}=&-i[\mathcal{H}_{MF}^\prime,\rho]+\kappa\left\{\sum_i (n_{\mathrm{th}}+1)L[a_i]\rho+n_{\mathrm{th}}L[a_i^\dagger]\rho\right\}\nonumber\\
&+\Gamma\sum_i L[\sigma_i^-]\rho,\label{11}
\end{align} 
in terms of the average thermal phonon number $n_{\mathrm{th}}=\left[\exp{(\hbar\omega/k_B T)}-1\right]^{-1}$. From this, we construct the free energy, $F=E-TS=\mathrm{Tr}(\rho\mathcal{H}_{MF}^\prime)+k_B T \mathrm{Tr}(\rho\ln{\rho})$ in terms of the average energy $E$, the entropy $S$, and the temperature $T$; where $k_B$ is the Boltzmann constant. The mean-field order paramater $\psi$ is obtained by minimizing the free energy $F$ in consistent with the constraint $\mathrm{Tr}(\rho)=1$. Driving and detuning conditions are kept same as in the case of zero temperature. The second-order coherence function $g^{(2)}(0)$ for temperatures of $T=1\mathrm{K}$, $5\mathrm{K}$, and $10\mathrm{K}$ is shown in Fig.~\ref{tempcoherence}a. The smooth transition of $g^{(2)}$ from phonon anti-bunching to bunching with increasing hopping $t$ are persistent upto temperatures of few Kelvins ($\sim 1\mathrm{K}$) for the non-equilibrium system; much higher than the predicted values for the equilibrium system with no driving field and dissipation present (see Fig.~\ref{Mottlobes}d). Entropy of the overall system is also calculated for a range of temperatures (Fig.~\ref{tempcoherence}b). For $T=1\mathrm{K}$, the entropy also approaches to zero with increasing hopping meaning that a pure superfluid phase is attained possessing the lowest possible energy that a quantum mechanical system can have.

\section{Summary}

We have considered the properties of arrays of strongly coupled cavity phonon-impurity two level systems in silicon, and we showed that small arrays will demonstrate new behavior and are realizable and measurable with present techniques. The observation of QPTs in large arrays will likely require extremely low effective temperatures (at least within the approximation considered here,\cite{Bradlyn2009}) i.e., for P:Si, $T=2\mbox{--}3$ mK ($g=1$ GHz), and for B:Si, $T=40\mbox{--}60$ $\mu$K ($g=21$ MHz).  (Our temperature results are equally applicable to polariton arrays, making circuit-QED many-body systems equally difficult to realize.) However, for a driven nonequlibrium system, our calculations of the second-order coherence functions still exhibit MI-SF quantum phase transitions up to a few degrees Kelvin. This indicates that driving may be a promising tool for the demonstration of QPT in solid state for finite temperatures. The true nature of temperature in the phonon-qubit array system and the potential for active cooling are also subjects worthy of further consideration. Our proposed many-body systems with phonons can be developed further for the pursuit of quantum simulators\cite{Feynman,Nori} or mediators between different quantum components and potentially for new quantum devices.


\end{document}